# Reflections on a Penetration Theory of Turbulent Heat Transfer

Trinh, Khanh Tuoc

*K.T.Trinh@massey.ac.nz*

---

## Abstract

This paper discusses a modified penetration theory where the time scale for the momentum wall layer is based on the onset of ejections from the wall but the time scale for the thermal layer is based on the unsteady state diffusion of heat from the wall into the streamwise flow, not the renewal of heat from eddies penetrating the wall layer.

A method for the determination of the thickness of the thermal wall layer has been developed. The theory correlates adequately experimental data for both the thickness of the thermal wall layer and the Nusselt number.

## 1 Introduction

The turbulent transport of heat, mass and momentum near a wall has been investigated intensively for more than a century. Since Reynolds (1874) proposed a that a mathematical analogy exists between the transport of these three quantities, much effort has focused on the extension of solutions of momentum to solutions for heat and mass transport. In view of the analogy any discussion on heat transfer will largely apply to mass transfer. There are a number of excellent reviews on heat transfer (Sideman and Pinczewski, 1975, Churchill, 1996, Mathpati and Joshi, 2007,

Pletcher, 1988). I will thus not be reviewing the existing work but will only mention the concepts and papers relevant to the topic at hand.

Essentially three approaches have been taken: a purely empirical approach typified by the Colburn analogy (1933), an analytical approach based on the solution time-averaged transport equations using turbulence models and an analogy approach where the Reynolds analogy is coupled with some model for the wall layer. In this latest category we need to differentiate between  models based on steady time averaged profiles (Karman, 1930, Martinelli, 1947, Lyon, 1951, Metzner and Friend, 1958) and penetration models based on the unsteady state diffusion equations (Danckwerts, 1951, Harriott, 1962, Ruckenstein, 1968, Hughmark, 1968, McLeod and Ponton, 1977, Thomas and Fan, 1971, Loughlin et al., 1985, Fortuin et al., 1992, Hamersma and Fortuin, 2003, Kawase and Ulbrecht, 1983).

Most studies start with a definition of the transport flux (e.g. of heat) in terms of a diffusive and a turbulent (also called eddy) contribution (Boussinesq, 1877)

$$q = -\left(k + \rho E_h\right)\frac{d\theta}{dy} \qquad (1)$$

| | | |
|---|---|---|
| where | $\theta$ | is the temperature |
| | y | the normal distance from the wall |
| | k | the thermal conductivity |
| | $E_h$ | the eddy thermal diffusivity |
| | q | the rate of heat transfer flux |
| | $\rho$ | the fluid density |

Equation (1) may be rearranged as

$$\theta^+ = \int_0^{y^+} \frac{q/q_w}{1 + E_h/k}\, dy^+ \qquad (2)$$

which is very similar to the equation for momentum transport

$$U^+ = \int_0^{y^+} \frac{\tau/\tau_w}{1 + E_\nu/\nu}\, dy^+ \qquad (3)$$

Where $U^+ = U/u_*$, $y^+ = yu_*\rho/\mu = yu_*/\nu$ and $\theta^+ = \rho C_p(\theta - \theta_w)u_*/q_w$ have been normalised with the friction velocity $u_* = \sqrt{\tau_w/\rho}$ and the fluid apparent viscosity $\mu$. The suffices w refer to the parameters at the wall, $\nu$ momentum, h heat and $\tau$, $E_\nu$ are the shear stress and eddy diffusivity for momentum respectively.

In the analytical approach a model is proposed for the term $E_h$. Usually it involves a number of parameters which are assigned empirically.

Reynolds' analogy states that

$$St = \frac{f}{2} \tag{4}$$

where

$St = \dfrac{h}{\rho C_p V}$ is called the Stanton number, (5)

$f = \dfrac{2\tau_w}{\rho V^2}$ the friction factor,

$C_p$ is the thermal capacity and

$V$ the average fluid velocity

This requires that the normalised velocity and temperature profiles be the same (Bird et al., 1960) p382.

$$\frac{d\theta^+}{dy^+} = \frac{dU^+}{dy^+} \tag{6}$$

It is normally assumed that Reynolds' analogy implies two conditions

$$q/q_w = \tau/\tau_w \tag{7}$$

$$\Pr_t = E_h/E_\nu = 1 \tag{8}$$

where $\Pr_t$ is called the turbulent Prandtl number. Equations (77) and (88) are at odds with experimental evidence. These are the paradoxes of the Reynolds analogy. The distribution of heat flux and shear stresses are not equal (Hinze, 1959, Churchill and Balzhiser, 1959, Seban and Shimazaki, 1951, Sleicher and Tribus, 1957) and the turbulent Prandtl number is not unity (Blom and deVries (Blom and deVries, 1968, McEligot et al., 1976, Malhotra and Kang, 1984, Kays, 1994, McEligot and Taylor, 1996, Weigand et al., 1997, Churchill, 2002). But if we move away from the wall into the log-law and outer region then we can obtain equation (6) from (1) and (2) by applying the following simplifications (Trinh, 1969)

$$\frac{E_h}{E_v} = \frac{q/q_w}{\tau/\tau_w}, \quad \frac{E_h}{k} >> 1 \quad and \quad \frac{E_v}{\nu} >> 1 \tag{9}$$

and equation (6) in Reynolds' analogy applies exactly outside a wall layer where the diffusive term predominates. The modelling of the diffusion layer in terms of pseudo-steady-state laminar flow (Karman, 1930, Martinelli, 1947, Lyon, 1951, Metzner and Friend, 1958, Reichardt, 1961, Levich, 1962) has received less attention by researchers in the last twenty years; many authors have switched to computer modelling investigate the velocity and temperature fields in greater details.

Penetration theories originated with Higbie (1935) who used the equation for unsteady conduction to model the transport process in jets and packed columns.

$$\frac{\partial \theta}{\partial t} = \alpha \frac{\partial^2 \theta}{\partial y^2} \tag{10}$$

where $\alpha = \rho C_P / k$ is the thermal diffusivity. The well-known solution is

$$q_w = \frac{1}{\sqrt{\pi}} \frac{\Delta\theta_m}{\sqrt{\alpha t_h}} \tag{11}$$

where $\Delta\theta_m = \theta_w - \theta_h$ is the maximum temperature drop between the wall and the edge of the laminar boundary layer and $t_h$ is a time scale for the diffusion process. Higbie closed the derivation by assuming that the typical time scale over which equation (11) applies is the contact time is

$$t_c = \frac{x}{U_\infty} \tag{12}$$

where x is the swept length. In an effort to apply Higbie's approach to turbulent transport, Danckwerts (1951) assumed that the surface near the wall is periodically swept clean by eddies penetrating from the bulk stream. The rate of renewal of the surface fluid near the wall is a function of the probability of occurrence of eddies of various frequencies. Danckwerts assumed this probability distribution to be uniform. Subsequent postulates of the surface renewal distributions have been reviewed by (Mathpati & Joshi, 2007; Pletcher, 1988; Ruckenstein, 1987; Sideman & Pinczewski, 1975). Many of these postulates do not link the assumed distribution of eddies to the improved understanding of the coherent structures or the wall structure but more recent work does e.g. Fortuin et al. (1992).

Ruckenstein (1968) first attempted to derive a physical model for the distribution function by modelling the eddy as a roll cell which circulates the fluid from the wall to the outer region. The motion close to the wall surface is assumed to obey the laminar transport equation

$$U\frac{\partial \theta}{\partial x} + V\frac{\partial \theta}{\partial y} = \alpha \frac{\partial^2 \theta}{\partial y^2} \tag{13}$$

Ruckenstein calls this state "pseudo-laminar flow" but does not elaborate about the relation between this state and the bursting phenomenon at the wall. Thomas and Fan (1971) used an eddy cell model proposed by Lamont and Scott (1970) in conjunction with a wall model by Black (1969) and the time scale measured by Meek and Baer (1970) to model the whole process. In both these approaches, the differentiation between the instantaneous fluxes and their time-averaged values is unclear and rough approximations are necessary to effect closure of the solution. Experimental measurements to vindicate these visualisations are difficult to obtain because the wall layer in mass transfer processes is extremely thin. Perhaps the most extensive studies have been attempted by Hanratty and his associates. Their ideas have evolved, along with improved experimental evidence, from a belief that the eddy diffusivity near the wall is proportional to $y^4$ at very high Schmidt numbers (Son and Hanratty, 1967), as predicted by Deissler (1955) to a belief that a more accurate power index is 3.38 (Shaw and Hanratty, 1964, 1977) to an argument that the analogy between heat and mass transfer breaks down completely very close to the wall (Na and Hanratty, 2000). The research of Hanratty showed that the characteristic length scale of mass transfer in the longitudinal direction is equal to that for momentum transfer (Shaw and Hanratty 1964, 1977) but the time scale for mass transfer is much shorter that for momentum transfer.

To explain this perplexing effect, Campbell and Hanratty (1983) have solved the unsteady mass transfer equations without neglecting the normal component of the convection velocity, which they model as a function of both time and distance. They found that only the low frequency components of the velocity fluctuations affect the mass transfer rates and that the energetic frequencies associated with the bursting process have no effect. In their explanation, the concentration sub-boundary layer acts as a low pass filter for the effect of velocity fluctuations on the mass transport close to the wall.

The existence of two time scales in the wall region of heat or mass transfer has been noted by all modern investigators. Their explanation is varied. McLeod and Ponton (1977) differentiate between the renewal period and the transit time which is defined as the average time that an eddy takes to pass over a fixed observer at the wall. Loughlin et al. (1985) and more recently Fortuin et al. (1992) differentiate between the renewal time and the age of an eddy. However, the exact identification of the renewal time with particular physical phenomena observed in turbulent flows remains vague. This is why penetration theories are still regarded with scepticism. For example Mathpathi and Joshi (op. cit.) stated that “these models serve a very limited purpose, because of the limited understanding of the relationships among the model parameters (contact time, renewal rate, size and shape of fluid packet, penetration depth of surface renewing eddies, etc.), and the flow parameters”.

This paper discusses the relationship between the modern observations of the wall process in turbulent flows and the parameters of a physically realistic penetration theory.

## 2 Theory

### 2.1 Equivalence of the penetration and boundary layer approaches to heat transfer

The first paper that I published when I was allowed to leave Viet Nam was meant to lay the foundation for a penetration theory of turbulent transport. It dealt with a transformation of the unsteady state conduction equation into the classical laminar boundary layer solutions (Trinh and Keey, 1992b). For simplicity we deal with transfer of heat from a flat plate to a Newtonian fluid stream but the arguments hold equally well for other geometries, fluid rheological characteristics and diffused quantities.

The governing equation for unsteady flow pat a flat plate with a zero pressure gradient is

$$\frac{\partial u}{\partial t}+u\frac{\partial u}{\partial x}+v\frac{\partial u}{\partial y}=\nu\frac{\partial^2 u}{\partial y^2} \tag{14}$$

Stokes neglected the convection terms to obtain

$$\frac{\partial u}{\partial t}=\nu\frac{\partial^2 u}{\partial y^2} \tag{15}$$

which was to apply to a flat plate suddenly set in motion. Similarly the governing equation for heat transfer is

$$\rho C_P\left(\frac{\partial\theta}{\partial t}+u\frac{\partial\theta}{\partial x}+v\frac{\partial\theta}{\partial y}+w\frac{\partial\theta}{\partial z}\right)=k\nabla^2\theta \tag{16}$$

For a laminar boundary layer past a flat plate, which can again be simplified to equation (10) by dropping the convection terms.

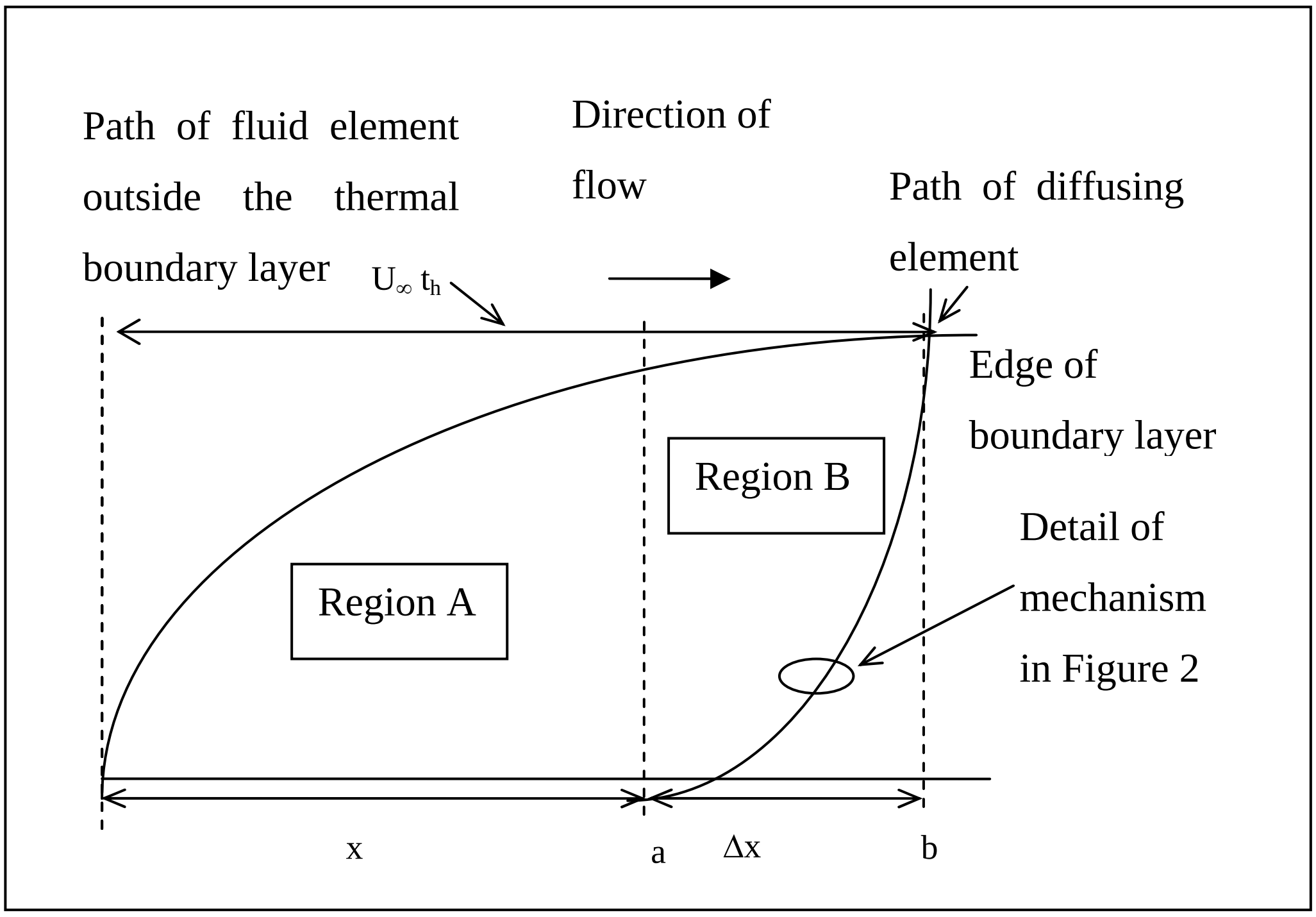

Figure 1 Diffusion path and convection velocities of an entity of heat after Trinh and Keey (1992).

We argued (Trinh and Keey, 1992b) that even in a steady state laminar boundary layer, heat and momentum continuously penetrate the fluid stream from the wall and furthermore each elemental packet of heat, called henceforth an entity of heat for convenience, travels across the boundary layer only once. Hence the appearance of steady state profiles can be seen as the result of an endless repetition of unsteady state movements of elements as shown in Figure 1.

Consider next the nature of forces acting on an entity of heat. At time t, the entity of heat enters an element of fluid at position (x, y), drawn in full line and coloured red in Figure 2, which has velocities, u and v. At time (t + δt), the element of fluid has moved by convection to a new position, not shown in Figure 2, and the thermal entity has diffused to an adjacent element at (x + δx, y + δy), drawn in dotted lines and coloured orange with a brick pattern. *Since the thermal entity is a scalar property, it has no mass*, feels the effect of diffusion forces only and convects with the velocity of the fluid element where it temporarily resides. *The convective forces act on the host element of fluid, not on the entity of heat.*

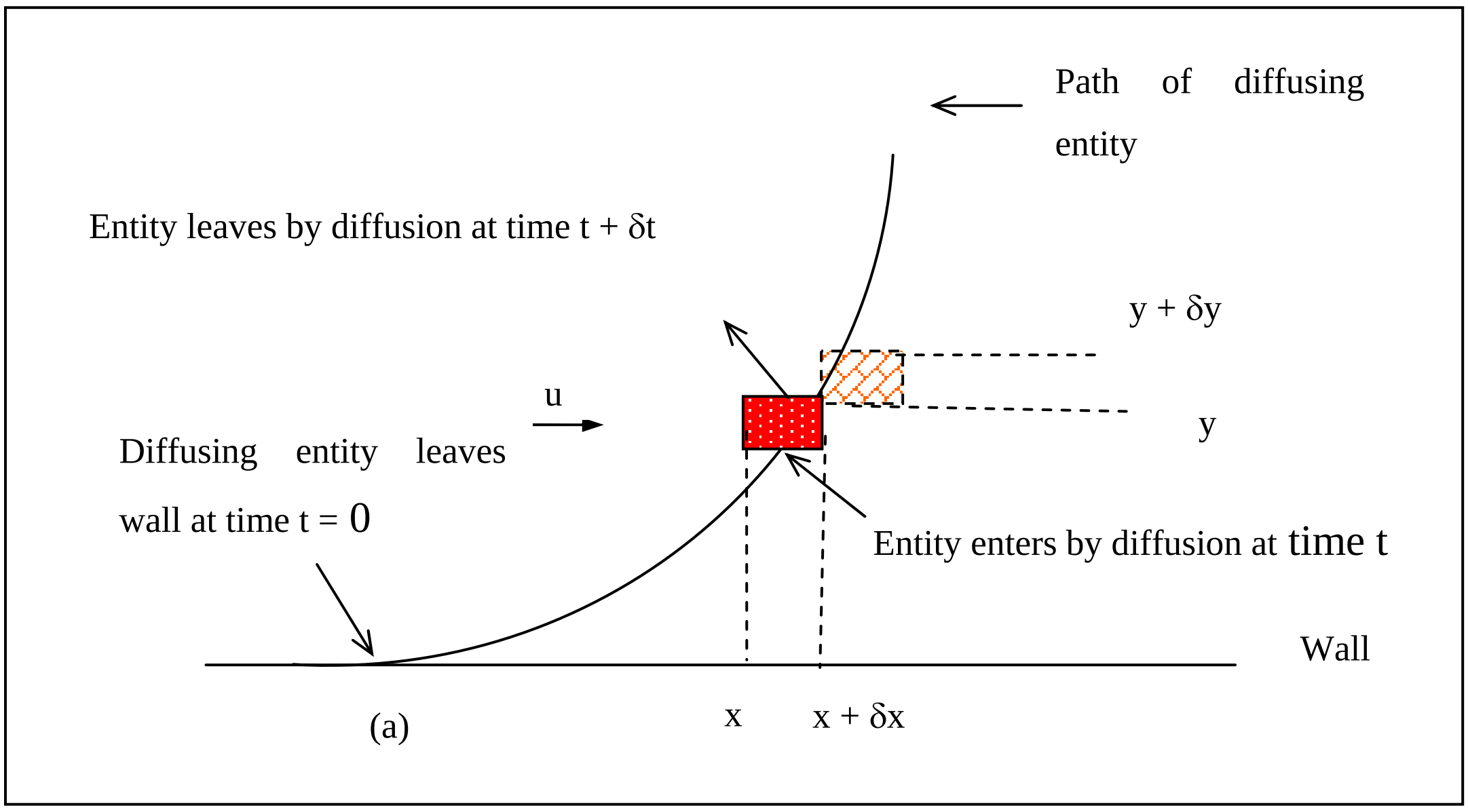

Figure 2. Convection and diffusion of an entity of heat

This physical analysis shows that the entities of fluid and heat move in different directions owing to different driving forces and it should be possible to separate the effects of diffusion and convection in the mathematical analysis. Consider now the mathematical description of this physical visualisation.

Equation (16) may be rewritten as

$$\frac{\mathrm{D}\theta}{\mathrm{Dt}} = \alpha \frac{\mathrm{D}^2\theta}{\mathrm{Dy}^2} \qquad (17)$$

where

$$\frac{D\theta}{Dt} = \frac{\partial\theta}{\partial t} = u\frac{\partial\theta}{\partial x} + v\frac{\partial\theta}{\partial y} + w\frac{\partial\theta}{\partial z} \qquad (18)$$

is called the substantial derivative.

Bird, Stewart, & Lightfoot (1960) p.73 illustrate the difference between the Eulerian partial derivative $\partial\theta/\partial t$ and the Lagrangian substantial derivative $D\theta/Dt$ with the following example. Suppose you want to count the fish population in a river. The Eulerian partial derivative gives the rate of change in fish concentration at a fixed point (x,y) in the river as seen by an observer standing on the shore. An observer in a boat drifting with the current will see the change in fish concentration on the side of the boat as given by the substantial derivative.

However, a fish swimming in the river will have a different perception of the fish population, which not described by either of these derivatives. Clearly a Lagrangian derivative is required but the convection velocities u and v are no longer relevant to this case; the velocity and path of the fish are. An observer attached to the *entity of heat* moving across the boundary layer will perceive the changes in temperature according to an equation similar to equation (10) but the frame of reference in the penetration theory is not attached to the wall as implied by Higbie (1935).

A second concern lays in the neglect of the convection terms in equations (14) and (15). In a laminar boundary layer the terms $\partial\theta/\partial t$, $u\partial\theta/\partial x$ and $v\partial\theta/\partial y$ are of the same magnitude and the simplification seems unjustified. However the physical analysis presented above makes the decoupling of convective and diffusive forces in line with physical reality if we interpret equation (10) in terms of a *Lagrangian derivative along the path of diffusion* (Trinh, 2010a).

$$\frac{\mathscr{D}\theta}{\mathscr{D}t} = \alpha \frac{\mathscr{D}^2\theta}{\mathscr{D}y^2} \qquad (19)$$

Then applying Taylor's hypothesis

$$\partial x = u\partial t \qquad (20)$$

transforms the solution of equation (10) into the classic solutions for steady state boundary layer heat transfer from a flat plate. Figure 3 shows that the unsteady state Lagrangian solution (in red dots) matches exactly the Eulerian solution of Polhausen (grid points on coloured background).

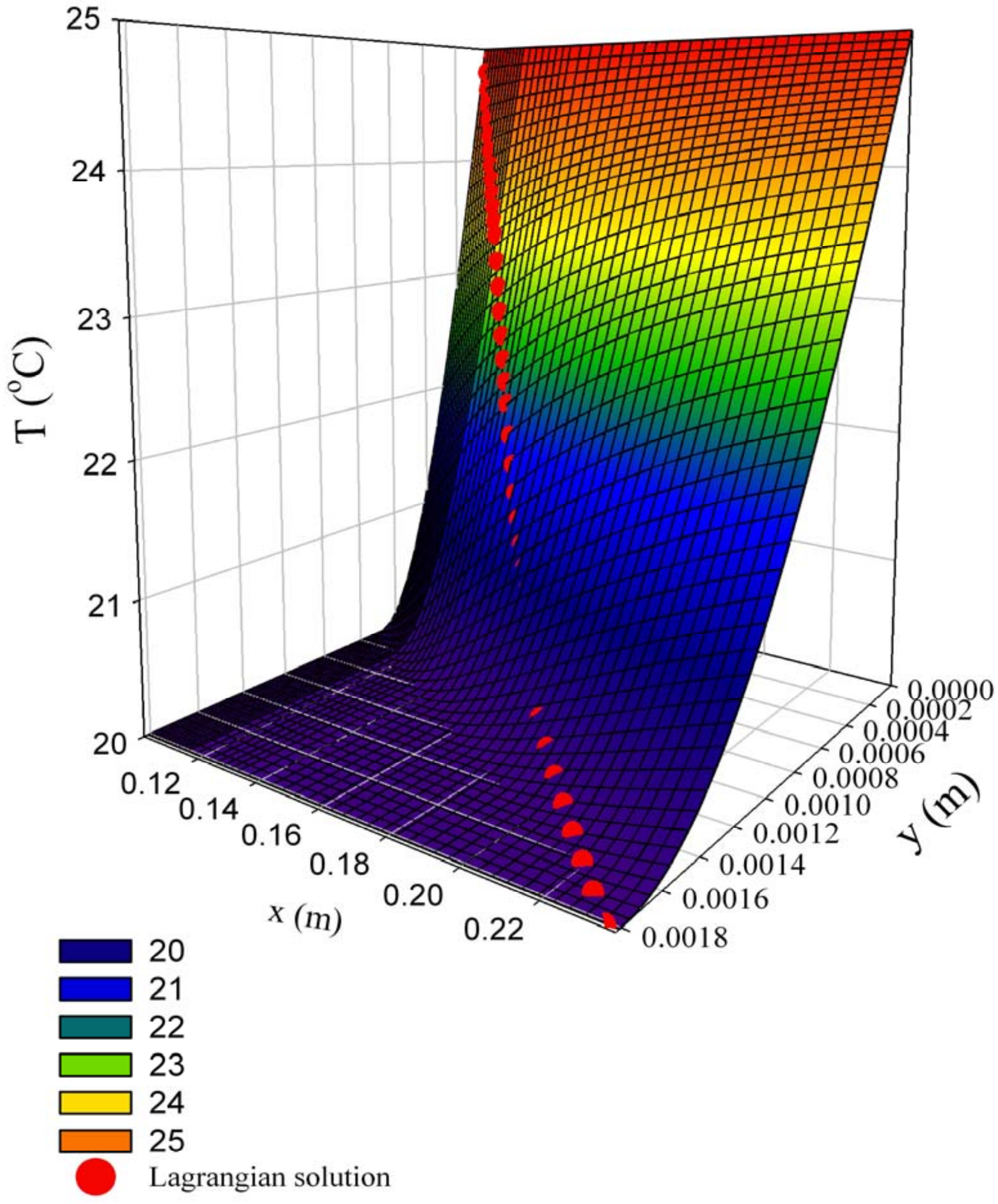

Figure 3 Diffusion path for a heat entity starting at the wall at x = 0.1 m and comparison with the Polhausen (1921) approximate solution for heat transfer in a laminar boundary layer.

We note that the entity of heat starting at the wall at location $x = 0.1m$ in this example reaches the edge of the thermal boundary layer at a position much further downstream ($x = 0.24m$) as indicated in Figure 1. Thus clearly the sets of matching pairs $(x, y)$ are not the same in the Eulerian Polhausen solution and in the Lagrangian solution.

Trinh and Keey (1992a) argued that the time scale of diffusion to be used in equation (19) must be linked with this distance $\Delta x$

$$t_h = \frac{\Delta x}{\bar{u}_B} \tag{21}$$

Where $\bar{u}_B$ is the average streamwise velocity in region B in Figure 1 and is related to the average velocity $\bar{u}_A$ in region A by

$$\frac{\bar{u}_A}{U_\infty} = 1 - \frac{\bar{u}_B}{U_\infty} = f(u) \tag{22}$$

Thus the time scale related to the entity of heat starting at the position (a) can be written as

$$t_h = \frac{x}{U_\infty f(u)} \tag{23}$$

Trinh and Keey showed that the function f(u) can be estimated by using the integral energy equation for boundary layer transport (Polhausen, 1921)

$$\frac{\partial}{\partial x} \int_0^{\delta_h} \frac{u}{U_\infty} \left( \frac{\theta - \theta_m}{\theta_w - \theta_m} \right) dy = \frac{h}{\rho C_p U_\infty} = St \tag{24}$$

with a third order polynomial temperature profile as

$$f(u) = \frac{8}{3} M \frac{U_\infty}{U_h} \tag{25}$$

where

$$M = \frac{\delta_h^*}{\delta_h} = \int_0^1 \frac{u}{U_\infty} \left( \frac{\theta - \theta_m}{\theta_w - \theta_m} \right) d\left( \frac{y}{\delta_h} \right) \tag{26}$$

is called usually a shape factor (Schlichting, 1979), p. 208.

$$\delta_h^* = \int_0^{\delta_h} \frac{u}{U_\infty} \left( \frac{\theta - \theta_m}{\theta_w - \theta_m} \right) dy \tag{27}$$

is the integral energy thickness, $\delta_h$ the thermal boundary layer thickness and $U_\infty$ and $U_h$ are the velocities at the edge of the viscous and thermal boundary layers respectively.

To obtain the parameter $M$ we need the velocity distribution. Trinh and Keey again used Polhausen third order polynomial model

$$\frac{u}{U_\infty} = 1.5\left(\frac{y}{\delta_v}\right) - \frac{1}{2}\left(\frac{y}{\delta_v}\right)^3 \tag{28}$$

Two situations must be distinguished depending on whether $U_h < U_\infty$ or $U_h > U_\infty$. The first situation corresponds to $\Pr >> 1$

Trinh and Keey showed that

$$M = \left(\frac{2}{5}\sigma - \frac{3}{280}\sigma^3\right) \tag{29}$$

where $\sigma = \delta_h / \delta_v = \Pr^{-b}$.

For $\Pr << 1$

$$M = \frac{3}{8} + \frac{1}{\sigma}\left(\frac{1}{4} - \frac{3}{5\sigma} + \frac{4}{35\sigma^3}\right) \tag{30}$$

$$f(u) = 1 + \frac{1}{\sigma}\left(\frac{2}{3} - \frac{8}{5\sigma} + \frac{32}{105\sigma^3}\right) \tag{31}$$

They obtain the parameter $b$ as a function of $\Pr$ as shown in Figure 4.

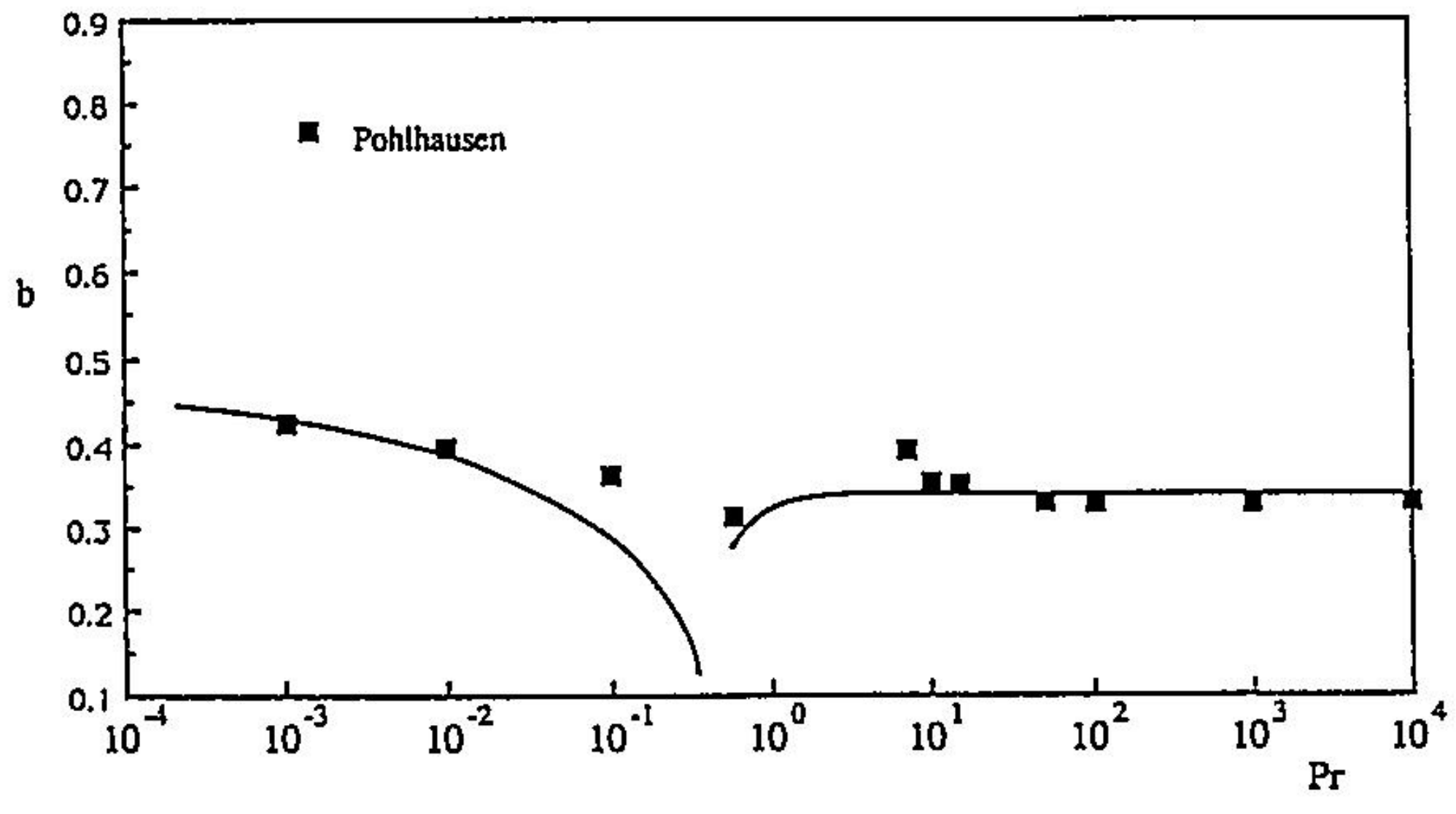

Figure 4 Exponent $b$ predicted by Trinh and Keey compared with Polhausen's (1921)

For $\Pr >> 1$, $b \approx 1/3$. Trinh and Keey showed that the Higbie time scale can be obtained by assuming plug flow $u = U_\infty$. The time scale for $\Pr >> 1$ is

$$t_h = \frac{x}{U_\infty\left(\frac{2}{5}\Pr^{-1/3} - \frac{1}{35}\Pr^{-1/27}\right)} \approx \frac{5x}{2U_\infty \Pr^{-1/3}} \tag{32}$$

For $\Pr = 1$, the thermal and momentum layer thicknesses are equal and

$$t_\nu = \frac{5x}{2U_\infty} \tag{33}$$

Thus the time scales for the thermal and momentum boundary layers are not equal

$$\frac{t_h}{t_\nu} = \frac{13\Pr}{14\Pr^{1-b} - 1} \approx \Pr^{1/3} \tag{34}$$

For $\Pr << 1$

$$\frac{t_h}{t_\nu} = \frac{39}{105 + 70\Pr^{b} - 168\Pr^{2b} + 32\Pr^{4b}} \tag{35}$$

## 2.2 Physical visualisation

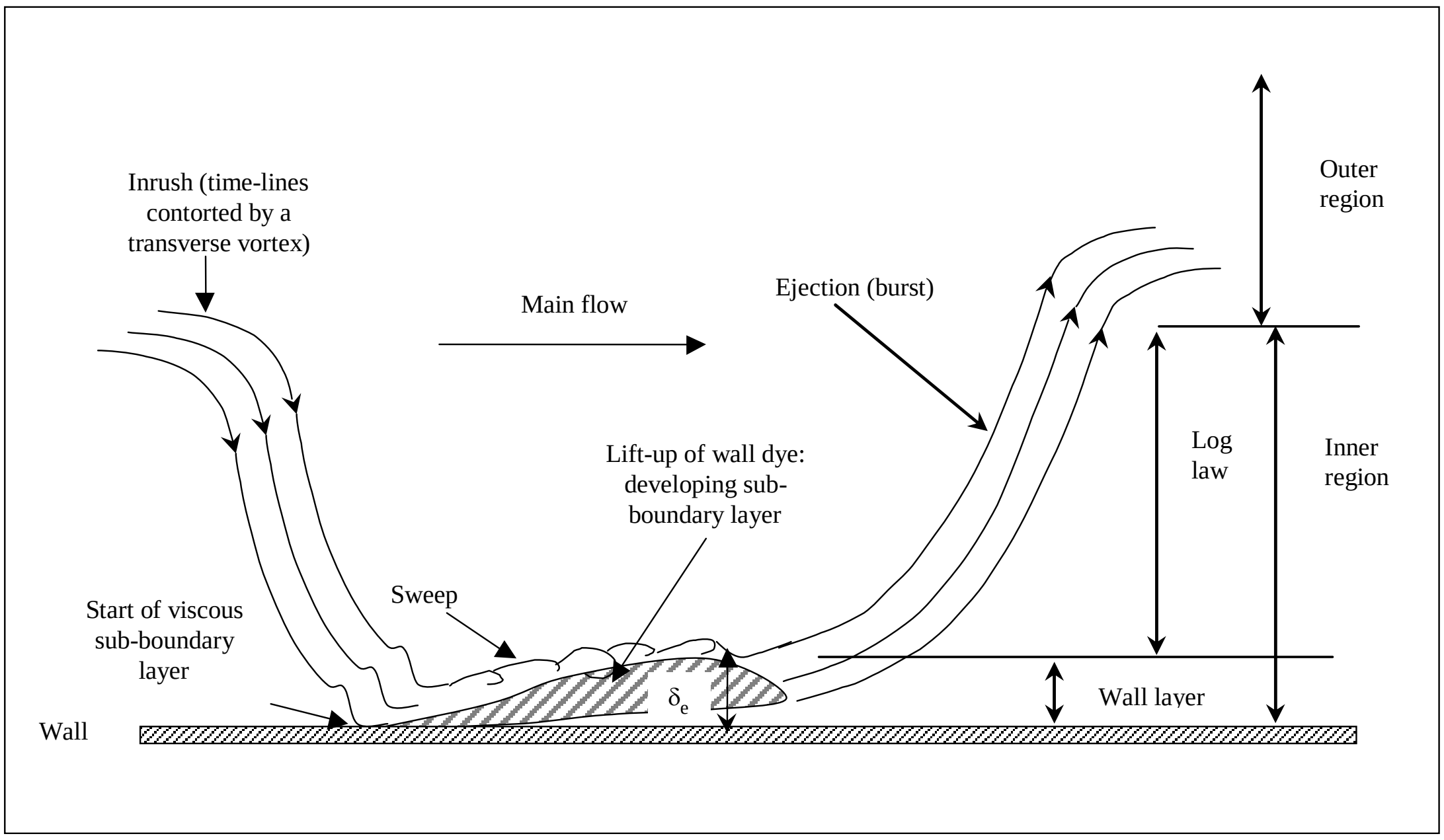

Figure 5. Schematic representation of the wall process in turbulent flow (Trinh, 2009).

The physical visualisation for turbulent transport underpinning this paper and all others in this theory of turbulence is based on the wall layer process first illustrated dramatically through hydrogen bubble tracers by Kline et al. (1967) and confirmed by many others. In plan view, Kline et al. observed a typical pattern of alternate low– and high-speed streaks. The low-speed streaks tended to lift, oscillate and eventually eject away from the wall in a violent burst.

In side view, they recorded periodic inrushes of fast fluid from the outer region towards the wall. This fluid was then deflected into a vortical sweep along the wall. The low-speed streaks appeared to be made up of fluid underneath the travelling vortex. The bursts can be compared to jets of fluid that penetrate into the main flow, and get slowly deflected until they become eventually aligned with the direction of the main flow.

The sweep phase, which lasts longest and dominates the statistics of the flow near the wall, can be modelled with the method of successive approximations borrowed from the analysis of oscillating laminar boundary layers (Schlichting, 1960, Tetlionis, 1981). The first approximation, called the solution of order $\varepsilon^0$, describes the diffusion of viscous momentum into the main stream. The solution of order $\varepsilon$ and higher only become important when the fast periodic velocity fluctuations have become strong enough to induce jets of fluid to be ejected from the wall i.e. during the bursting phase (Trinh, 2009).

In other words, the wall layer defined by the solution of order $\varepsilon^0$ is visualised as an unsteady state laminar sub-boundary layer which is interrupted by the emergence of the ejections. Mass, heat and momentum are contained in the same body of fluid ejected from the wall which explains, in my view, why there is an analogy between the laws of heat, mass and momentum in the outer region.

Since the low speed streaks occur randomly in time and space, a fixed probe will return a statistical average temperature and velocities of low-speed streaks of all ages. There are of course many other coherent structures in turbulent flows e.g. (Robinson, 1991) but they do not seem contribute the long term statistical averages and both the profiles of velocity (and temperature) and the transport fluxes can be adequately be modelled with the two structures discussed here: the low speed streaks and the ejections (Trinh, 2010c).

## 2.3 The paradox of time scales

Most authors have long recognised that the Higbie time scale is unable to correlate steady state convective heat transfer. The differentiation between the residence time of a packet of fluid at the wall and the frequency at which it is renewed by fluid from the bulk flow proposed in existing penetration theories does not explain, in my view, the difference between the two time scales in equations **Error! Reference source not found.**31) and **Error! Reference source not found.**32) because surely the momentum, heat and mass in the wall layer must be renewed simultaneously by the incoming eddies. For example, if one explains the second time scale in the thermal wall layer in terms of the "age" of eddies sweeping the wall (Fortuin et al, 1991) one is faced with the question as to why the age of ejected lumps of fluid from the wall does not affect Reynolds' analogy. Indeed any argument based on *convective* forces which are specific to heat and mass transfer (i.e. not present in momentum transfer) raises a paradox when one applies Reynolds' analogy.

In the present theory, there is only one mechanism for agitation in a turbulent boundary layer. This agitation process relies on the intermittent ejection of wall fluid into the outer region and its time scale is $t_\nu$, the time scale of the momentum wall layer. The second time scale $t_h$ reflects a *diffusion* process within the wall layer and therefore does not relate to the agitation mechanism; it is related to the diffusion of heat and mass across the wall layer and best understood if one analyses equations **Error! Reference source not found.**15) and **Error! Reference source not found.**19) in a Lagrangian context. Because of the difference in momentum and thermal diffusivities, the depth of penetration of heat and viscous momentum from the wall differ as shown by the conceptual illustration of a mapping of the velocity and temperature contours in Figure 6. Contours of velocity and temperature in the wall layer. Figure 6.

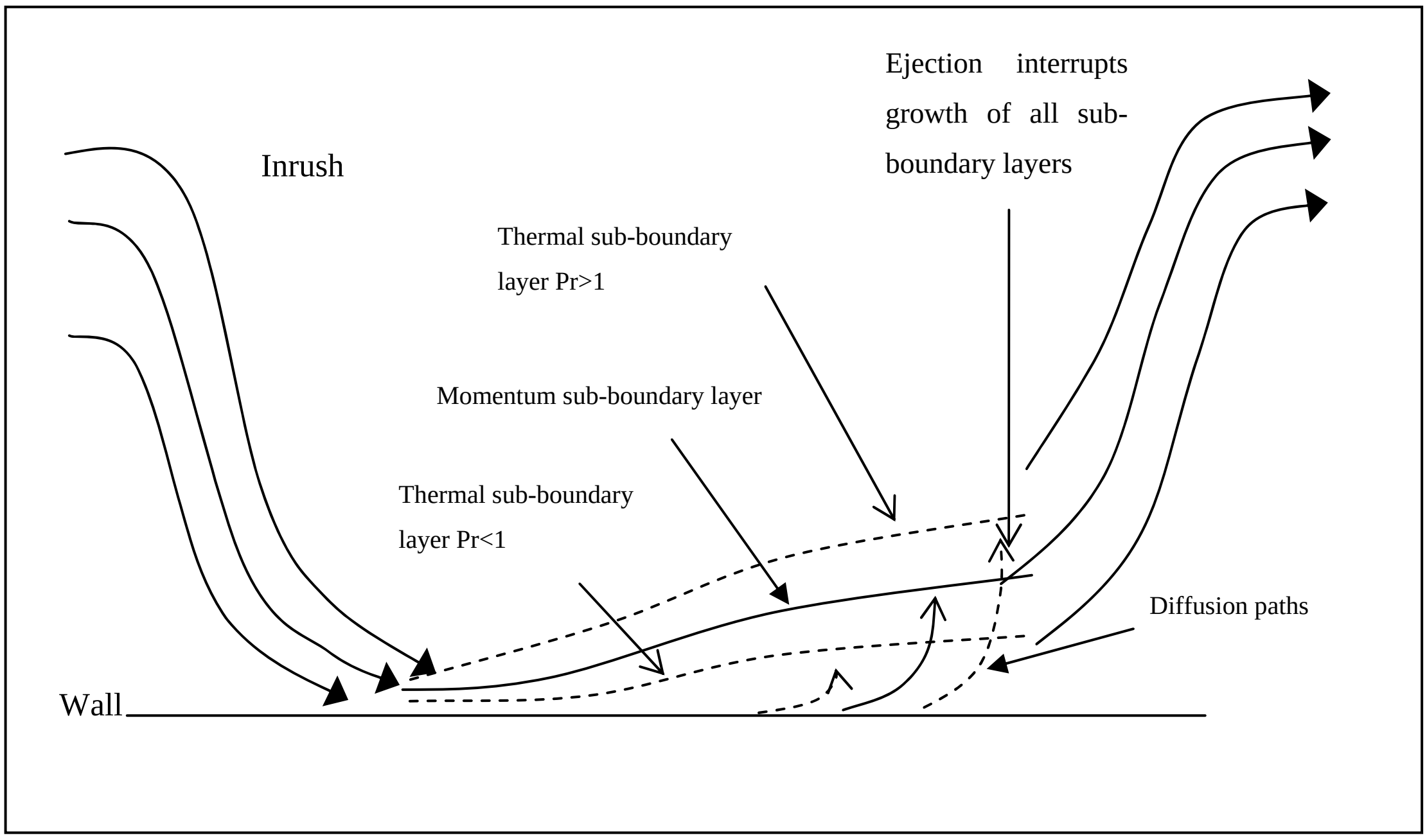

Figure 6. Contours of velocity and temperature in the wall layer.

In my view there is only one defining time scale for the wall layer $t_\nu$ because it defines the moment when the ejection occurs and therefore sets the life time of the low speed streak. Both the momentum and thermal content of the wall fluid are regenerated at the same time because of the subsequent inrush from the main stream. The surface renewal theory would require two different agitation events: the ejections and other separate rushes of fluid into the wall during the sweep phase. I have never seen any publication giving evidence of eddies or streams penetrating through the hair pin vortices into the low-speed streaks.

Most previous penetration theories postulate that turbulent eddies from the main flow penetrate into the wall layer to renew selectively its heat content without apparently affecting the momentum. The present visualisation argues that both heat and momentum penetrate from the wall into the main flow but at different rates.

## 2.4 Mathematical formulations

The equation for diffusion of viscous momentum

$$\frac{\mathscr{D}\tilde{u}}{\mathscr{D}t} = \nu \frac{\mathscr{D}^2\tilde{u}}{\mathscr{D}y^2} \qquad (36)$$

Can be solved by the method of Stokes (1851) for the conditions:

| | | | |
|---|---|---|---|
| IC | t = 0 | all y | $\tilde{u} = U_\nu$ |
| BC1 | $t > 0$ | y = 0 | u = 0 |
| BC2 | $t > 0$ | y = ∞ | $\tilde{u} = U_\nu$ |

Where $\tilde{u}$ is the smoothed phase velocity of the low speed streak. The well-known solution is

$$\frac{\tilde{\mathrm{u}}}{\mathrm{U}_\nu} = \mathrm{erf}\,(\eta_{\mathrm{s}}) \qquad (37)$$

where $\eta_s = \dfrac{y}{\sqrt{4\nu t}}$ (38)

The average wall-shear stress is

$$\tau_w = \frac{\mu U_\nu}{t_\nu}\int_0^{t_\nu}\left(\frac{\partial\tilde{u}}{\partial y}\right)_{y-0} dt = \frac{\mu U_\nu}{t_\nu\sqrt{\pi}}\int_0^{t_\nu}\frac{1}{\sqrt{\nu t}}dt \qquad (39)$$

Equation **Error! Reference source not found.** may be rearranged as

$$\mathrm{t}_\nu^+ = \frac{2}{\sqrt{\pi}}\mathrm{U}_\nu^+ \qquad (40)$$

The time-averaged velocity profile near the wall may be obtained by rearranging equation (37) as

$$\frac{U^+}{U_\nu^+} = \int_0^1 erf\left(\frac{y^+\sqrt{\pi}}{4U_\nu^+\sqrt{t/t_\nu}}\right)d\left(\frac{t}{t_\nu}\right) \qquad (41)$$

Equation (41) applies up to the edge of the wall layer where $\mathrm{u}/\mathrm{U}_\nu = 0.99$, which corresponds to $\mathrm{y} = \delta_\nu$ and $\eta_{\mathrm{s}} = 1.87$. Substituting these values into equation **Error! Reference source not found.** gives

$$\delta_\nu^+ = 3.74 t_\nu^+ \qquad (42)$$

where $t_\nu^+ = u_*\sqrt{t_\nu/\mu}$. Back-substitution of equation **Error! Reference source not found.** into **Error! Reference source not found.** gives

$$\delta_\nu^+ = 4.22 U_\nu^+ \qquad (43)$$

$$\frac{\theta - \theta_h}{\theta_w - \theta_h} = erf(\eta_h) \tag{44}$$

where $\eta_h = \frac{y}{\sqrt{4\alpha t}}$ (45)

which can be arranged as

$$\eta_h = \frac{y^+ \Pr^{1/2}}{2t^+} \tag{46}$$

The time average wall heat flux is

$$\begin{aligned} q_w &= \frac{k(\theta_w - \theta_m)}{t_h} \int_0^{t_h} \left( \frac{\partial(\theta_w - \theta)}{\partial y} \right)_{y=0} dt \\ q_w &= \frac{k(\theta_w - \theta_m)}{t_h \sqrt{\pi}} \int_0^{t_h} \frac{1}{\sqrt{\alpha t}} dt \\ q_w &= \frac{2}{\sqrt{\pi}} \frac{k(\theta_w - \theta_m)}{\sqrt{\alpha t_h}} \end{aligned} \tag{47}$$

Equation (47) may be rearranged as

$$t_h^+ = \frac{2}{\sqrt{\pi} \Pr} \theta_h^+ \tag{48}$$

The time-averaged temperature profile near the wall may be obtained by rearranging equation **Error! Reference source not found.** as

$$\frac{\theta^+}{\theta_h^+} = \int_0^1 erf\left( \frac{y^+ \Pr \sqrt{\pi}}{4\theta_h^+ \sqrt{t/t_h}} \right) d\left( \frac{t}{t_h} \right) \tag{49}$$

Equation (49) applies up to the edge of the wall layer where $u/U_\nu = 0.99$, which corresponds to $y = \delta_h$ and $\eta_h = 1.87$. Substituting these values into equation (46) gives

$$1.87 = \frac{\delta_h^+ \Pr^{1/2}}{2t_h^+} \tag{50}$$

Substituting equation (48) into (50) gives

$$\theta_h^+ = \frac{\delta_h^+ \Pr}{4.22} \tag{51}$$

There are two alternative estimates for the time scale $t_h$ that we will both investigate.

### 2.4.1 Different time scales for diffusion of momentum and heat

In this case, we follow the arguments of Trinh and Keey that the time scales for diffusion of heat and momentum in a boundary layer are different because the rates of diffusion are governed by coefficients $\alpha$ and $\nu$ that do not have the same value. We must again differentiate two cases

$\Pr > 1$

Combining equations (34), (42) and (50) gives

$$\delta_h^+ = \frac{\delta_\nu^+}{\Pr^{1/2}} \frac{t_h^+}{t_\nu^+} = \frac{\delta_\nu^+}{\Pr^{1/2}} \sqrt{\frac{13 \Pr}{14 \Pr^{1-b} - 1}} \tag{52}$$

Taking $b \approx 1/3$ (Trinh and Keey 1992a), then

$$\delta_h^+ = \frac{\delta_\nu^+}{\Pr^{1/2}} \sqrt{\frac{13 \Pr}{14 \Pr^{2/3} - 1}} \tag{53}$$

Now if we write for convenience

$$\delta_h^+ = \delta_\nu^+ \Pr^a \tag{54}$$

equation (53) can be used to show that for $\Pr > 1$, $a = -1/3$.

$\Pr < 1$

Combining equations (35), (42) and (50) gives

$$\delta_h^+ = \frac{\delta_\nu^+}{\Pr^{1/2}} \frac{t_h^+}{t_\nu^+} = \frac{\delta_\nu^+}{\Pr^{1/2}} \sqrt{\frac{34}{105 + 70 \Pr^b - 168 \Pr^{2b} + 32 \Pr^{4b}}} \tag{55}$$

where $b \to 1/2$ as $\Pr \to 0$ (Trinh and Keey, 1992a).

### 2.4.2 Same time scale for diffusion of heat and momentum

The argument here is that when the pocket of fluid at the wall is ejected it interrupts both before the diffusion of heat and momentum at the same time. Thus

$$t_h^+ = t_\nu^+ \tag{56}$$

Combining equations (42), (50) and (56)

$$\delta_h^+ = \frac{\delta_\nu^+}{\Pr^{1/2}} \tag{57}$$

## 3 Comparison with experimental measurements

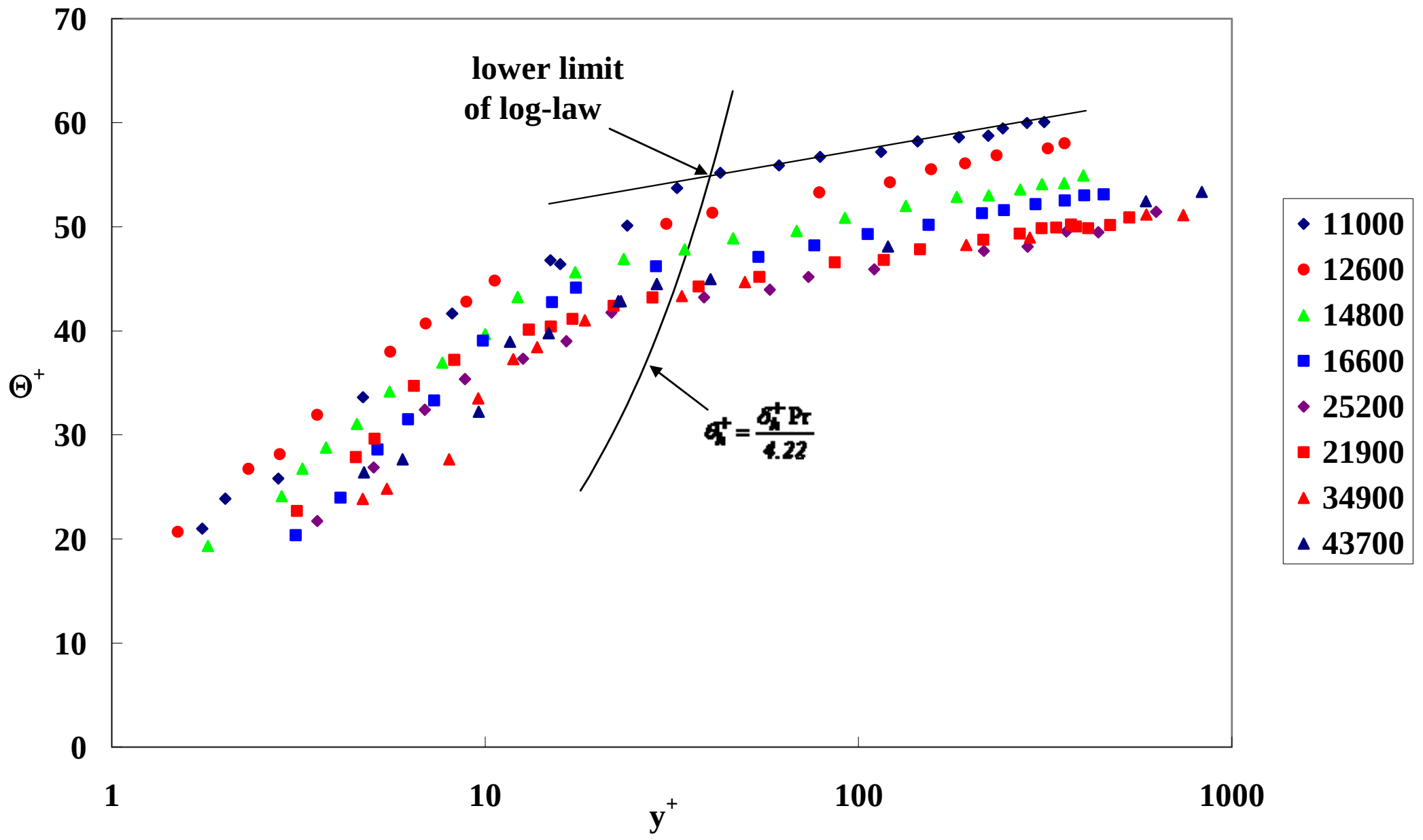

Figure 7. Temperature data of Smith et al. (1967) for $Pr = 5.7$ at different Reynolds number and determination of thermal wall layer thickness.Smith et al. indicate that the transition region for heat transfer extends to higher Reynolds numbers than for momentum transfer as observed by many authors e.g. (Gnielinski, 1976, Trinh et al., 2010)

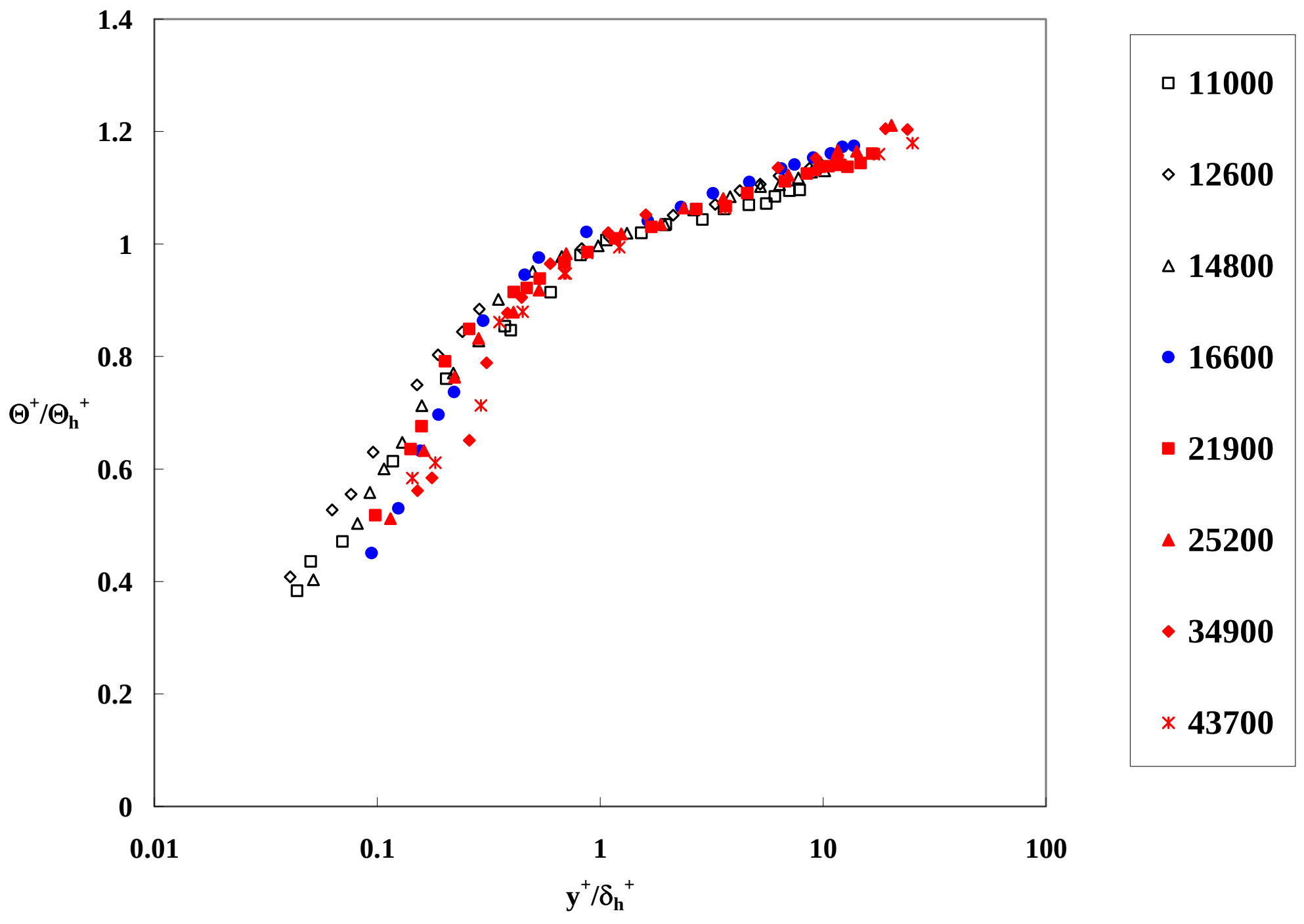

Figure 8. Zonal similarity profile of temperature. Same data as in Figure 7.

The temperature profiles become almost independent of Reynolds number above $Re > 20000$.

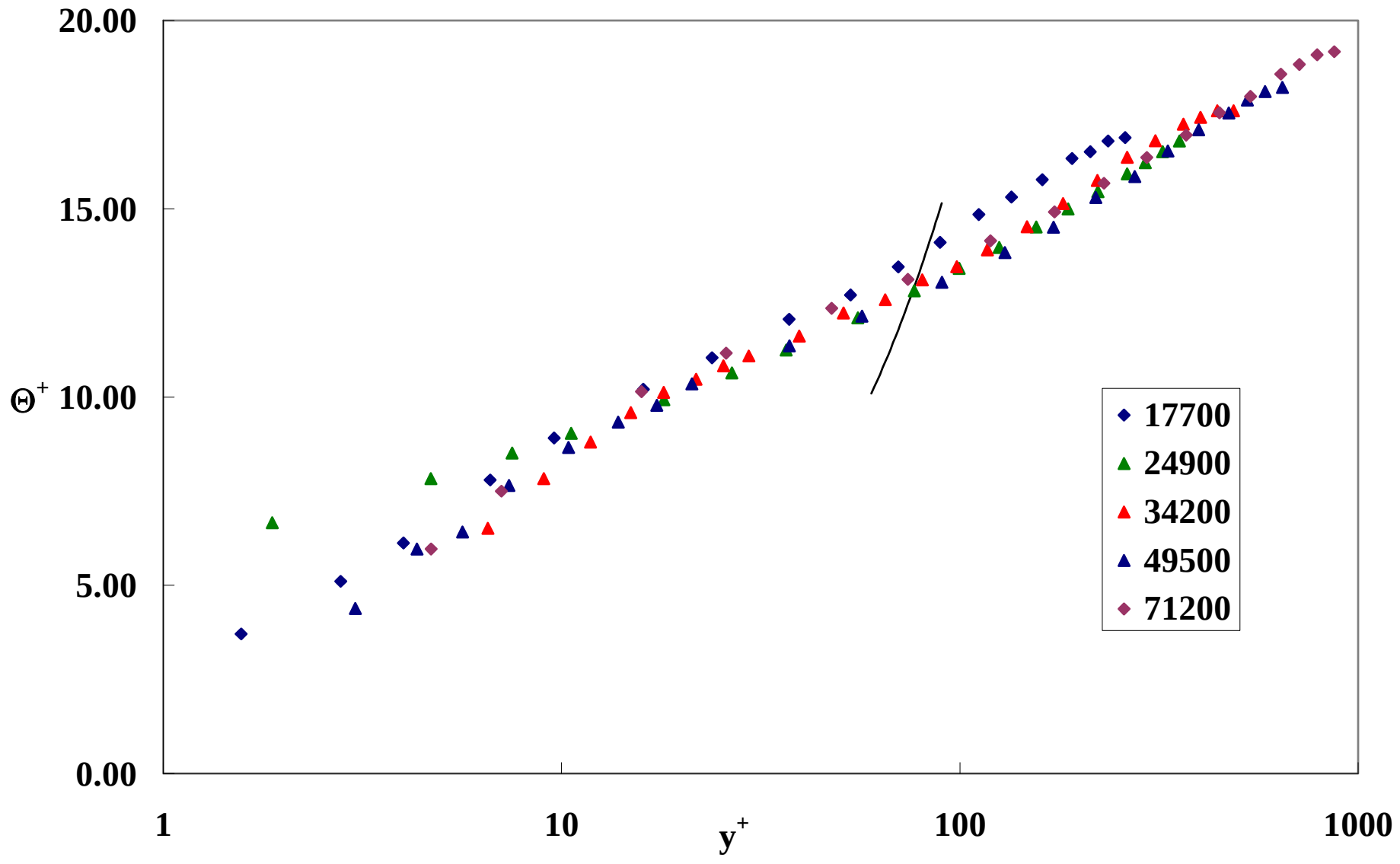

Figure 9. Temperature profiles for air (Johnk and Hanratty, 1962), $Pr = 0.7$.

Normalising the temperature and distance with their values at the edge of the wall layer removes the effect of the Reynolds number just as found for velocity profiles (Trinh, 2010d). Another example is shown for air with $Pr = 0.7$ in Figures 9 and 10.

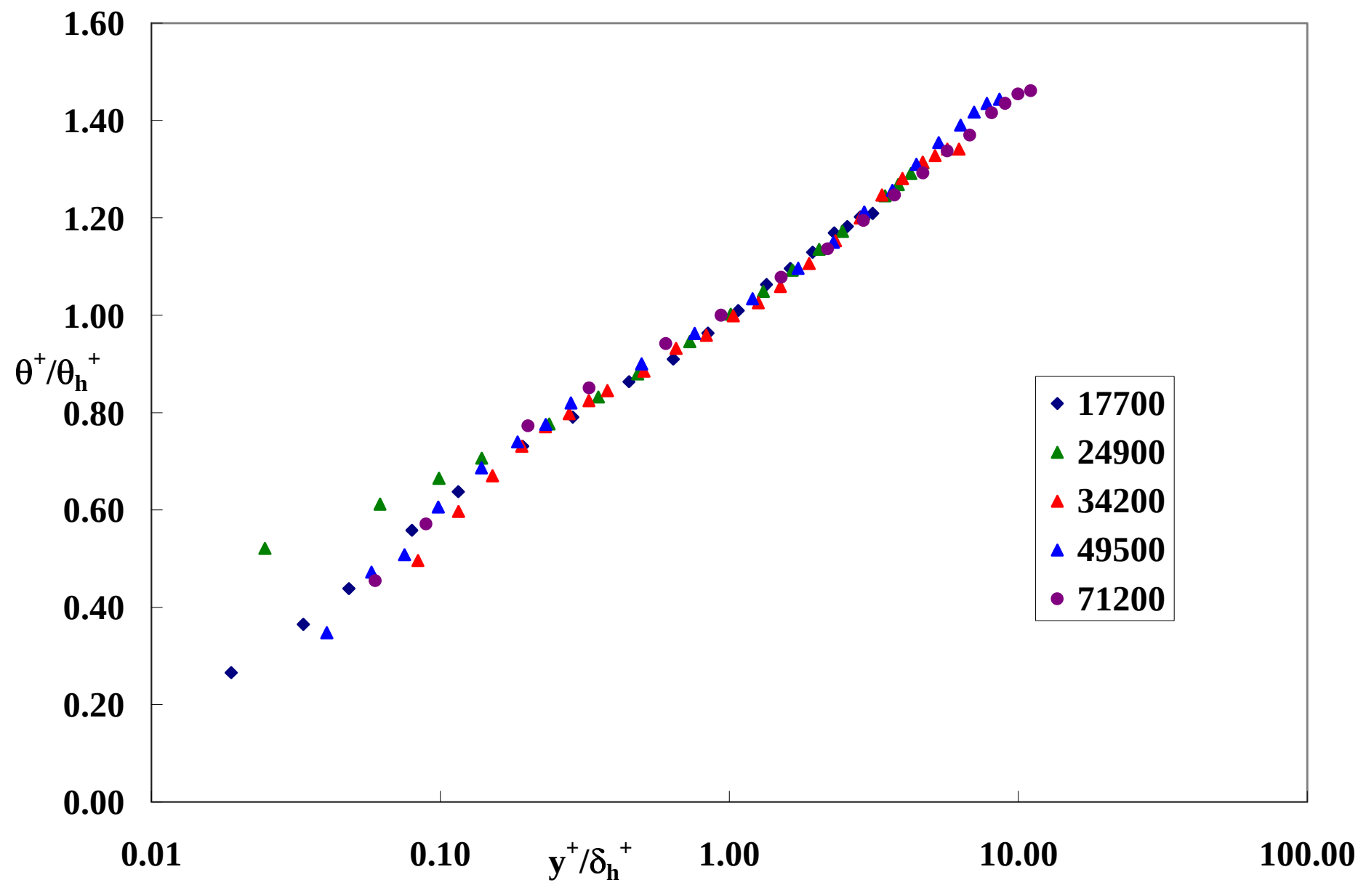

Figure 10. Zonal similarity temperature profiles for air. Same data as in Figure 9.

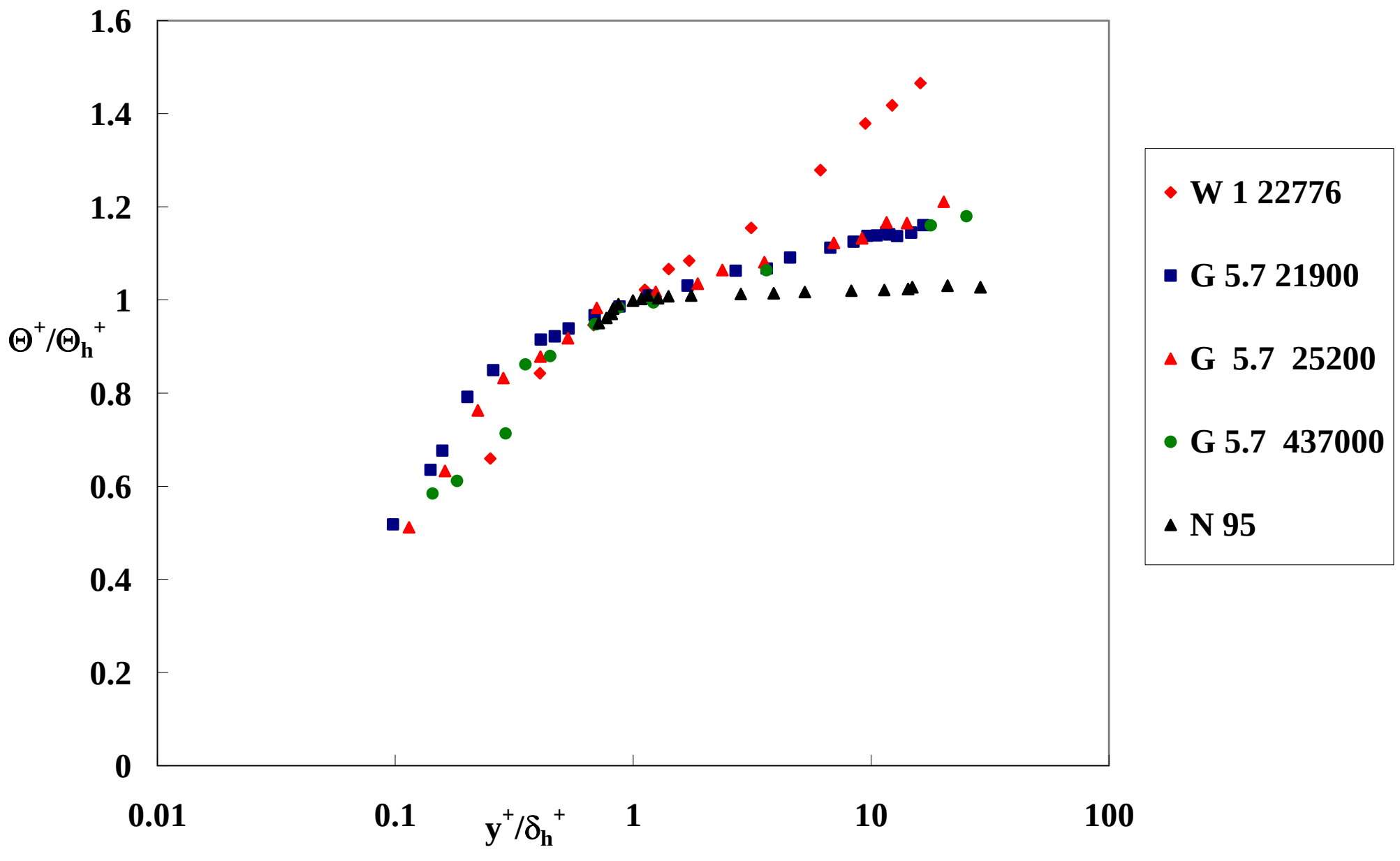

Figure 11. Zonal similarity plot of velocity profile W (Wei and Willmarth, 1989),

This inability to collapse the temperature profiles highlights the fact that the mechanism in the log-law layer is based on a convection principle, which is independent of the Prandtl number whereas the mechanism in the wall layer is based on a diffusion principle which is dependent of the Prandtl number as shown in section 2.4.

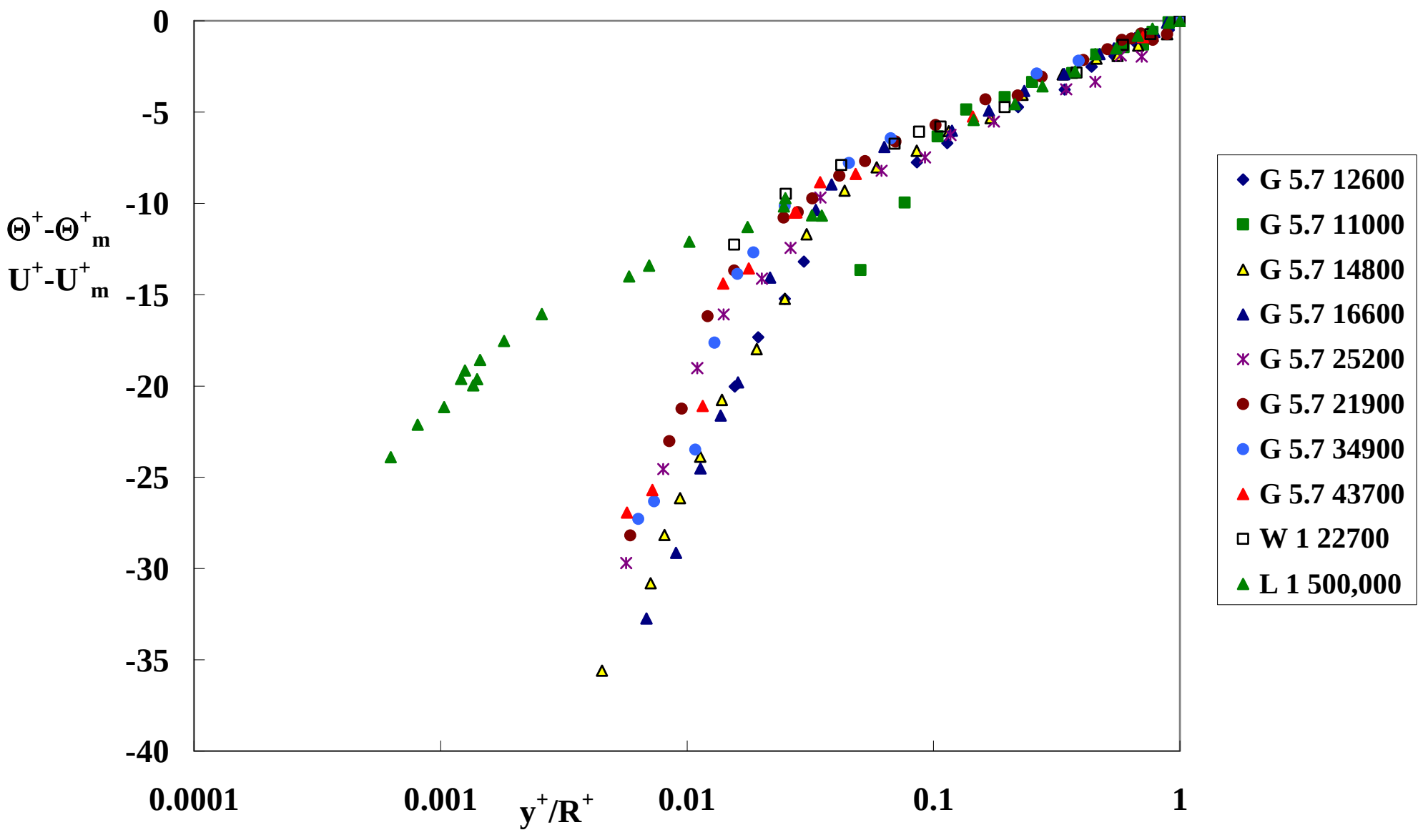

Figure 12. Velocity and temperature defect plot showing Reynolds analogy . Data G (Smith et al., 1967) Pr– *5.7* , W (Wei and Willmarth, 1989), L (Laufer, 1954) velocity,

However, Reynolds' analogy applies well in the outer region as shown in the velocity and temperature defect plot in Figure 14.

A plot of $\delta_h^+ / \mathrm{Pr}^a$ vs. Re where a is given by equations (54) and (55) collapses the heat and momentum transfer data as shown in Figure 13. In particular, the data support $a \approx 1/3$ for $\mathrm{Pr} \geq 5.7$. By contrast, normalisation of $\delta_h^+$ with $\mathrm{Pr}^{1/2}$ as suggested by equation (57) does not collapse the thermal and momentum thicknesses as shown in Figure 14.

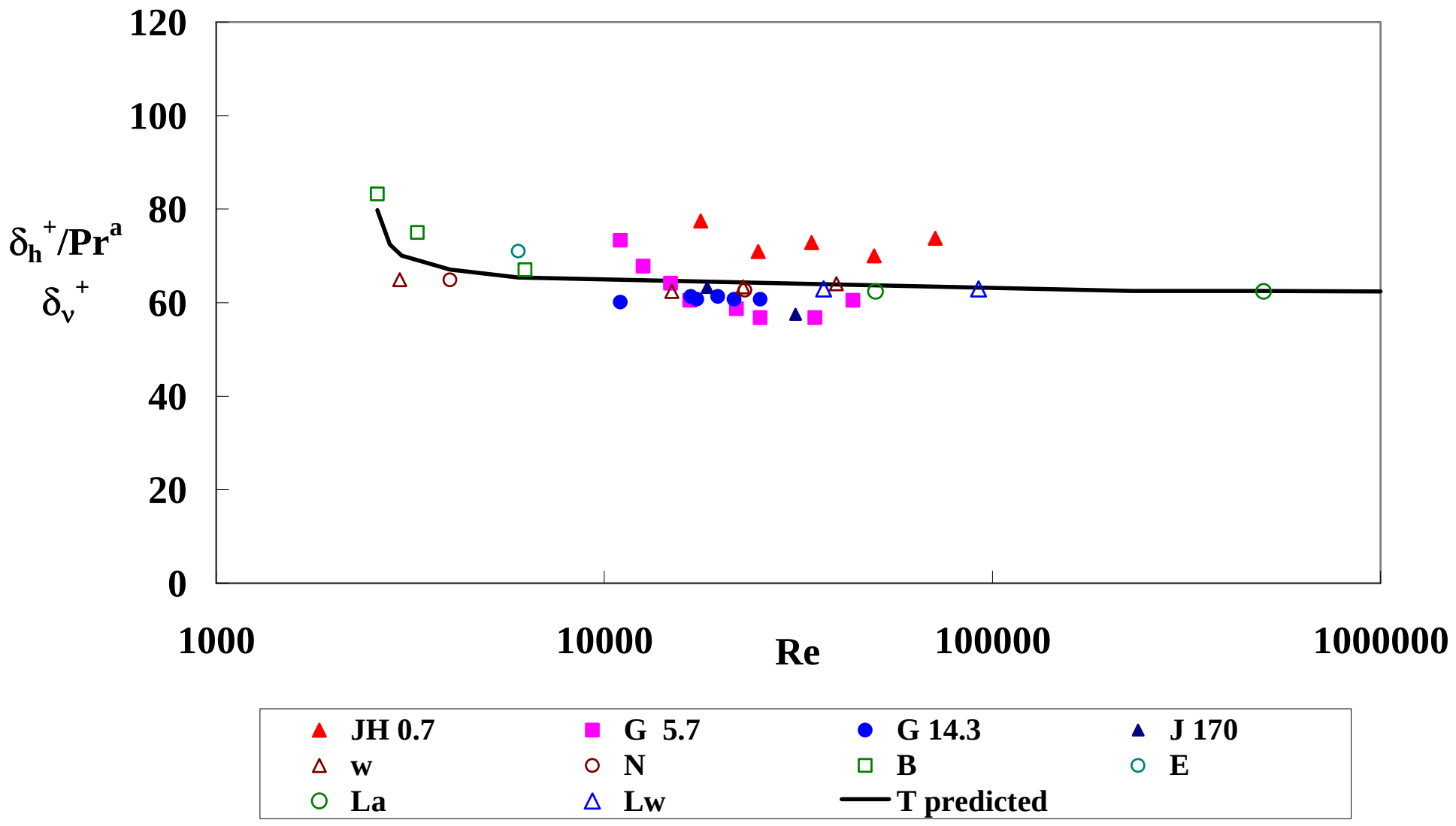

Figure 13. Variation of thermal wall layer thickness with Reynolds and Prandtl numbers. Index $a$ given by equations (54) and (55). Data JH (Johnk and Hanratty, 1962), G (Smith et al., 1967), J (Janberg, 1970), W (Wei and Willmarth, 1989), N (Nikuradse, 1932), B (Bogue, 1962), E (Eckelmann, 1974), La (Laufer, 1954), Lw (Lawn, 1971), T (Trinh, 2010b). Thermal data filled points. Velocity data hollow points.

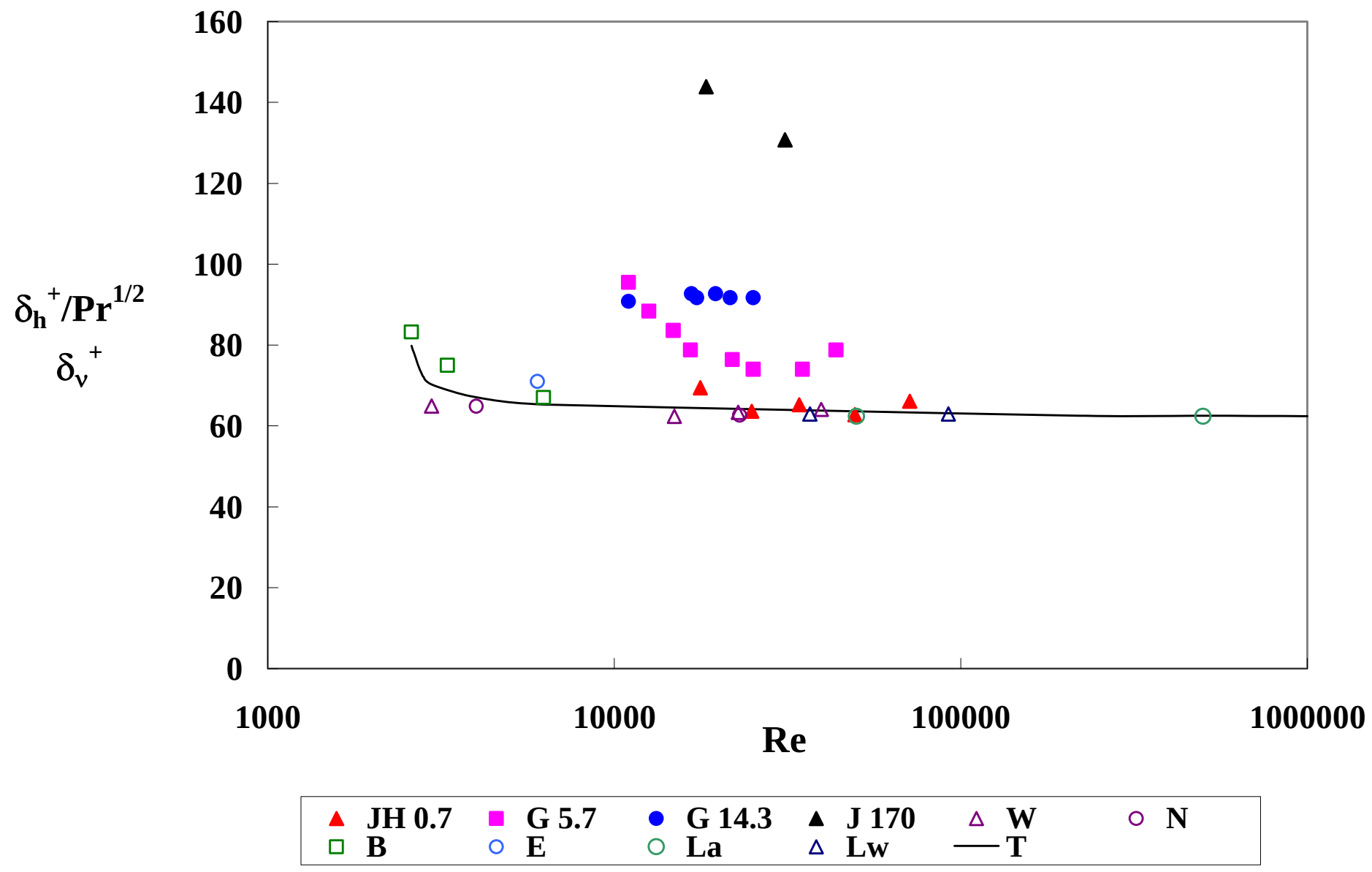

Figure 14. Variation of thermal wall layer thickness normalised with equation (57) against Reynolds and Prandtl numbers. Same data as in Figure 13.

Applying Reynolds' analogy in the outer region in the log-law region

$$\Theta^+ = 2.5 \ln\left(\frac{y^+}{\delta_h^+}\right) + \Theta_h^+ \qquad (58)$$

For $\Pr > 1$ we substitute $\delta_h^+ = \delta_v^+ \Pr^{-1/3}$ and $\Theta_h^+ = \delta_v^+ \Pr^{2/3} / 4.22$ into equation (58)

$$\Theta^+ = 2.5 \ln\left(\frac{y^+ \Pr^{1/3}}{\delta_v^+}\right) + \frac{\delta_v^+ \Pr^{2/3}}{4.22} \qquad (59)$$

For fully turbulent flow $\delta_v^+ = 64.8$ (Trinh, 2009) then

$$\Theta^+ = 2.5 \ln\left(y^+ \Pr^{1/3}\right) + 15.35 \Pr^{2/3} - 10.43 \qquad (60)$$

At the pipe axis, $y^+ = R^+$, $\Theta^+ = \Theta_m^+$ and $R^+ = \frac{Re}{2}\sqrt{\frac{f}{2}}$ and the Nusselt number can be derived from the temperature profile by standard techniques (e.g. Schlichting, 1960). Introducing

$$D(\Pr, Re) = \Theta_m^+ - \Theta_b^+ \qquad 61)$$

where $\Theta_b^+$ is the mixing cup temperature given by

$$\Theta_b^+ = \frac{1}{\pi R^{+2}} \int \Theta^+ U^+ 2\pi \left(R^+ - y^+\right) dy^+ \qquad (62)$$

The value of $D(\Pr, \mathrm{Re})$ has been calculated by performing the integration indicated by equation (62) numerically and tabulated (Trinh, 1969).

$$\Theta_b^+ = 2.5 \ln\left(\frac{\mathrm{Re}\sqrt{f}}{2\sqrt{2}}\Pr^{1/3}\right) + 15.35\Pr^{2/3} - 10.43 - D(\Pr, \mathrm{Re}) = \frac{\sqrt{f/2}}{St} \quad (63)$$

$$St = \frac{\sqrt{f/2}}{2.5\ln \mathrm{Re}\sqrt{f} + 15.55\Pr^{2/3} - 13.03 + (2.5/3)Ln\Pr - D(\Pr, \mathrm{Re})} \quad (64)$$

Equation (64) is shown against the data of Friend (1958) who measured both the Nusselt number ( $Nu = St * \mathrm{Re} * \Pr$ ) and friction factor in Figure 15. It has the same level of accuracies as other analogy formulae derived in this theory of turbulence (Trinh, 2009). These will be discussed in a summarising review of heat transfer correlations.

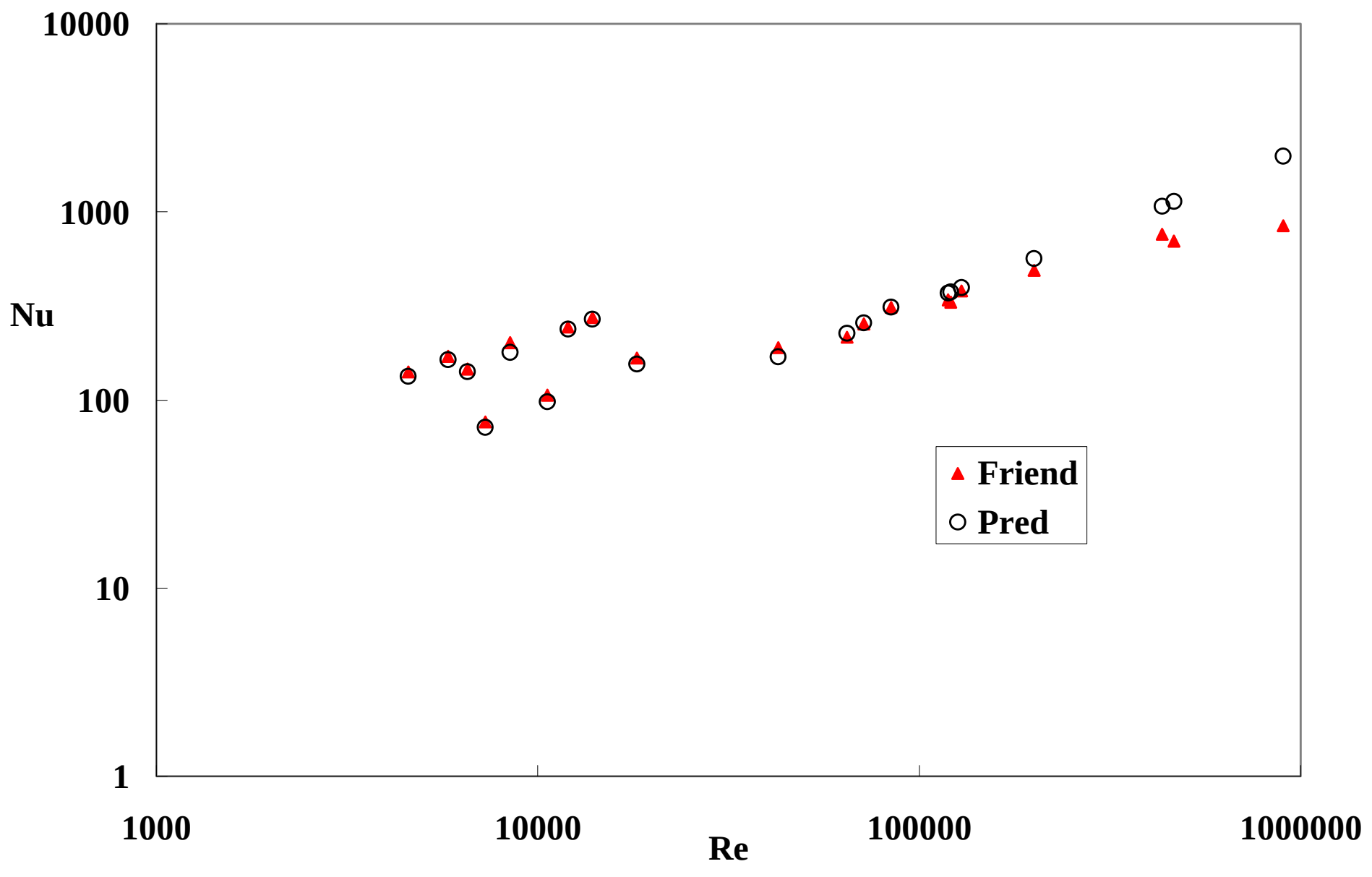

Figure 15. Comparison of equation (64) against data of Friend (1958) $1.88 \le \Pr \le 264$.

# 4 Conclusion

A method for determination the thickness of the thermal wall layer has been developed. A penetration theory is discussed where the time scale for the momentum wall layer is based on the onset of ejections from the wall but the time scale for the thermal layer is based on the unsteady state diffusion of heat from the wall into the streamwise flow, not the renewal of heat from eddies penetrating the wall layer. The thickness of the thermal wall layer is well correlated by this technique. Reynolds' analogy is confirmed in the region beyond the wall layer. Predictions for the Nusselt number also correlate well with experimental data. In contrast the predictions based on a unique time scale for both momentum and thermal transfer are found not accurate.